\documentclass[a4paper,fleqn,usenatbib,letters]{mnras}

\usepackage[T1]{fontenc}
\usepackage{ae,aecompl}

\usepackage{graphicx}	
\usepackage{amsmath}	
\usepackage{amssymb}	
\usepackage{lscape}
\usepackage{tikz}   
\usepackage{gensymb}
\usepackage{tikz-3dplot} 
\usepackage{import}
\usepackage{lmodern}
\RequirePackage{fix-cm}

\newcommand{\nobs}{\mathbf{n}_{\rm obs}}

\newcommand{\ey}{\mathbf{e}_{\rm y}}
\newcommand{\np}{\mathbf{n}_{\rm p}}
\newcommand{\npi}{\mathbf{n}_\Pi}
\newcommand{\rlight}{r_{\rm L}}

\title[Off-centred dipole polarization]{Polarized emission from of an off-centred dipole.}

\author[J. P\'etri]{J.  P\'etri
\thanks{E-mail: jerome.petri@astro.unistra.fr} \\
  Observatoire astronomique de Strasbourg, Universit\'e de Strasbourg, CNRS, UMR 7550, 11 rue de l'universit\'e, 67000 Strasbourg, France.
  }

\date{Accepted XXX. Received YYY; in original form ZZZ}

\pubyear{2016}

\begin{document}
\label{firstpage}
\pagerange{\pageref{firstpage}--\pageref{lastpage}}
\maketitle

\begin{abstract}
Radio polarization measurements of pulsed emission from pulsars offer a valuable insight into the basic geometry of the neutron star: inclination angle between the magnetic and rotation axis and inclination of the line of sight. So far, all studies about radio polarization focused on the standard rotating vector model with the underlying assumption of a centred dipole. In this letter, we extend this model to the most general off-centred dipole configuration and give an exact closed analytic expression for the phase-resolved polarization angle. It is shown that contrary to the rotating vector model, for an off-centred dipole, the polarization angle also depends on the emission altitude. Although the fitting parameter space increases from two to six (position of the dipole, altitude and shift of the zero phase), statistical analysis should remain tractable. Observations revealing an evolution of the polarization angle with frequency would undeniably furnish a strong hint for the presence of a decentred magnetic dipole in neutron stars.  
\end{abstract}




\section{Introduction}

The wealth of radio pulse profile and polarization data acquired about pulsar emission over half a century still faces some problems to be satisfactorily and synthetically explained by a unique geometrical model. The extensively used rotating vector model (RVM) introduced by \citet{radhakrishnan_magnetic_1969} and \cite{komesaroff_possible_1970} to explain the evolution in the polarization angle (P.A.) of radio pulsars represents an interesting phenomenological approach to account for those observations. Their interpretation is based on the simple fact that curvature radiation of charged particles along magnetic field lines is polarized in the plane of curvature. Invoking purely geometrical considerations, they showed that this polarization angle~$\psi$ measured with respect to the projection of the stellar rotation axis onto the plane of the sky is given for a centred magnetic dipole by
\begin{equation}
\label{eq:RVM}
 \tan \psi = \frac{\sin\chi \, \sin \varphi}{\sin\zeta \, \cos \chi - \cos\zeta \, \sin \chi \, \cos\varphi}
\end{equation}
with $\varphi=\Omega\,t$ the pulse phase, $\chi$ the obliquity and $\zeta$ the inclination of the line of sight, the origin of phase being chosen appropriately. The stellar rotation rate is~$\Omega$ and the time of observation is~$t$. The variation in the P.A.~$\psi$ is maximum for zero longitude~$\varphi=0$ and equals
\begin{equation}
 \left( \frac{d\psi}{d\varphi} \right)_{\rm max} = \frac{\sin\chi}{\sin(\zeta-\chi)} .
\end{equation}
Although very simple, this model reproduces the characteristic S-shaped curves and some other nice features summarized by \cite{chung_stokes_2011} who used Stokes parameters to investigate polarization. However the RVM has recently been hampered for twenty pulsars \citep{yan_polarization_2011} and it is well known that the S-shaped swing is challenged by other millisecond pulsars \citep{xilouris_characteristics_1998}. The limitation of this phenomenological approach becomes apparent. Moreover, the RVM ignores the rotation of the star, especially aberration effects, delay and swept back field lines. These effects are not relevant for slow pulsars because corrections scale as the ratio $R/\rlight$ \citep{phillips_radio_1992} where $R$ is the neutron star radius and $\rlight$ the light cylinder radius. The impact of relativistic effects on polarization was discussed by \cite{blaskiewicz_relativistic_1991}. The inflection point of the linear polarization angle no longer coincides with the peak of the radio intensity but arrives out of phase approximately with an estimate of $4\,r/\rlight$, where $r$ represents the height of the emission region. Moreover, including plasma effects \cite{hibschman_polarization_2001} added the deformation of field lines by magnetospheric current. \cite{blaskiewicz_relativistic_1991} approximation breaks down above roughly~$0.1\,\rlight$ thus at too low altitudes for millisecond pulsars. This encouraged \cite{craig_altitude_2012} to improve it by more accurately describing finite height corrections for higher altitudes. \cite{craig_tackling_2014} gave good quantitative fits by allowing also for orthogonal mode jumps.

In addition to the RVM, emission regions shaped as concentric cones were introduced to explain the diversity of pulse profiles: single, double or multiple components. Invoking one or more cones and a variable inclination of the line of sight, almost any type of profile is reproducible. \cite{komesaroff_linear_1970} based his model on the hollow cone \citep{radhakrishnan_magnetic_1969}, an argument also taken up by \cite{backer_pulsar_1976} to explain the behaviour of pulse profiles from radio emission. In this model, emission is the result of curvature radiation in curvilinear motion along field lines emanating from the polar caps. Curvature radiation being impossible along straight lines, curvature photons cannot come from inside the cone, resulting in an absence of emission in the vicinity of the polar axis, justifying the hollow cone model. However, to explain the variety of observed pulses, Backer had to add to his model a central component. The central component is wider and has a softer spectrum \citep{lyne_shape_1988}. The hollow cone model was criticized by \cite{izvekova_applicability_1977} and is not unanimously approved. \cite{smith_beaming_1970} had already allocated the radio emission to a source in rotation with the star and situated in the vicinity of the light cylinder, severely contrasting with near surface emission expectations.


Does the star contain a perfectly centred dipole as assumed by the RVM? Not really actually. Observational support for an off-centred dipole has been found for main sequence stars. Indeed \cite{stift_decentred_1974} looked at a decentred dipole in stars with a displacement vector along the magnetic axis. Such dipoles are useful to solve the asymmetry problem between the north and south hemisphere as explained by \cite{landstreet_orientation_1970}. An off-centred dipole is also the preferred way to interpret Zeeman line profiles \citep{borra_orientation_1974}. For some pulsars like PSR~J2144-3933 the radio emission is explained with a novel pair creation model in the magnetosphere \citep{zhang_radio_2000} or simply by the presence of intense multipolar components of the surface magnetic field in all radio pulsars \citep{gil_vacuum_2001, gil_modelling_2002}. Multipolar components are a natural consequence of a rotating off-centred dipole \citep{petri_radiation_2016}.

We note that despite the accumulation of data on radio pulsations, no model or beginning of explanation has the favour of observers. In my view, increasing the receiver sensitivity to detect even more pulsars will not help us much. Twenty five years ago, in 1992, \cite{radhakrishnan_polarization_1992} already said , ``more and more detailed radio observations are NOT what is needed to help with theoretical modelling.'' The answer will maybe come from a multi-wavelength modelling. Indeed, most pulsars are detected in radio only, but some of them also emit in other wavelengths of the electromagnetic spectrum, from radio waves up to gamma-rays through optical and X-rays. The Crab pulsar is a prototypical case where pulsed emission is also observed in infra-red, optical, X-rays and gamma-rays.

In this letter, we generalize the RVM by allowing for a translation between the centre of the star and the location of the magnetic moment, we call it decentred RVM (DRVM). The exact geometry is layed out in Sec.~\ref{sec:Modele} and extends the RVM model given by eq.~(\ref{eq:RVM}). In Sec.~\ref{sec:Resultats} we briefly expose the main feature of our DRVM. Discussion of the results and future plans are given in Sec.~\ref{sec:Discussion}.

\section{Rotating off-centred dipole}
\label{sec:Modele}

For the parameters of an off-centred dipole, we follow the notations introduced by \cite{petri_radiation_2016} for a radiating dipole in vacuum. Here we neglect retardation effects and rotational sweep back of magnetic field lines. First we recall the important geometrical quantities and the magnetic configuration. Second we compute exactly and analytically the polarization angle for any rotational phase $\varphi=\Omega\,t$ and any geometry. Vectors are expanded onto Cartesian orthonormal basis.

\subsection{Geometrical set-up}

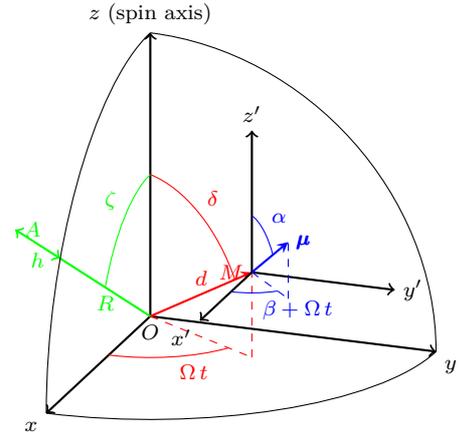
\begin{figure}
\centering


\tdplotsetmaincoords{70}{110}

\pgfmathsetmacro{\rvec}{.6}
\pgfmathsetmacro{\thetavec}{60}
\pgfmathsetmacro{\phivec}{60}

\begin{tikzpicture}[scale=4,tdplot_main_coords]

\coordinate (O) at (0,0,0);

\tdplotsetcoord{P}{\rvec}{\thetavec}{\phivec}

\tdplotsetcoord{A}{1}{60}{0}

\tdplotsetcoord{B}{1.5}{60}{0}


\draw[thick,->] (0,0,0) node[below] {$O$} -- (1,0,0) node[anchor=north east]{$x$};
\draw[thick,->] (0,0,0) -- (0,1,0) node[anchor=north west]{$y$};
\draw[thick,->] (0,0,0) -- (0,0,1) node[anchor=south]{$z$ (spin axis)};

\tdplotsetrotatedcoords{0}{0}{0}

\draw[-stealth,thick,color=red] (O) -- (P) node [midway, above] {$d$} node [above, left] {$M$} ;
\draw[dashed,color=red,tdplot_rotated_coords] (0,0,0) -- (0.26,0.45,0);
\draw[dashed,color=red,tdplot_rotated_coords] (0.26,0.45,0) -- (0.26,0.45,.3);
\tdplotdrawarc[tdplot_rotated_coords,color=red]{(0,0,0)}{0.4}{0}{60}{anchor=north west}{$\Omega\,t$}

\draw[thick,color=green] (O) -- (A) node [midway, below] {$R$} ;
\draw[<->,thick,color=green] (A) -- (B) node [midway, below] {$h$} node [above, right] {$A$} ;



\tdplotsetthetaplanecoords{\phivec}

\tdplotdrawarc[red, tdplot_rotated_coords]{(0,0,0)}{0.5}{0}{\thetavec}{anchor=south west}{$\delta$}

\tdplotsetthetaplanecoords{0}

\tdplotdrawarc[green, tdplot_rotated_coords]{(0,0,0)}{0.5}{0}{60}{anchor=south east}{$\zeta$}


\tdplotsetrotatedcoords{0}{0}{0}

\tdplotsetrotatedcoordsorigin{(P)}

\draw[thick,tdplot_rotated_coords,->] (0,0,0) -- (.5,0,0) node[anchor=north east]{$x'$};
\draw[thick,tdplot_rotated_coords,->] (0,0,0) -- (0,.5,0) node[anchor=west]{$y'$};
\draw[thick,tdplot_rotated_coords,->] (0,0,0) -- (0,0,.5) node[anchor=south]{$z'$};


\draw[-stealth,thick,color=blue,tdplot_rotated_coords] (0,0,0) -- (.2,.2,.2) node [right] {$\pmb{\mu}$} ;
\draw[dashed,color=blue,tdplot_rotated_coords] (0,0,0) -- (.2,.2,0);
\draw[dashed,color=blue,tdplot_rotated_coords] (.2,.2,0) -- (.2,.2,.2);

\tdplotdrawarc[tdplot_rotated_coords,color=blue]{(0,0,0)}{0.2}{0}{45}{anchor=north west}{$\beta+\Omega\,t$}

\tdplotsetrotatedthetaplanecoords{45}

\tdplotdrawarc[tdplot_rotated_coords,color=blue]{(0,0,0)}{0.2}{0}{55}{anchor=south west}{$\alpha$}

\begin{scope}[canvas is xy plane at z=0]
     \draw (1,0) arc (0:90:1);
\end{scope}
\begin{scope}[canvas is xz plane at y=0]
     \draw (1,0) arc (0:90:1);
\end{scope}
\begin{scope}[canvas is yz plane at x=0]
     \draw (1,0) arc (0:90:1);
\end{scope}
   
\end{tikzpicture}
\caption{Geometry of the decentred dipole showing the three important angles $\{\alpha, \beta, \delta\}$ and the distance $d$. Two additional parameters related to observations are the line of sight inclination~$\zeta$ and the emission height~$h$. The plot corresponds to time~$t$ assuming that $\bmu$ lies in the $(xOz)$ plane at $t=0$.}
\label{fig:Dipole}
\end{figure}

The neutron star is depicted as a solid body in uniform rotation at a rate~$\Omega$. Its magnetic moment is located inside the sphere of radius~$R$ at a point~$M$ such that at any time $t$ its position vector is
\begin{equation}
\mathbf{d} = d \, (\sin \delta \, \cos \Omega\,t, \sin \delta \, \sin \Omega\,t, \cos \delta)
\end{equation}
where $d$ is the distance from the centre and $\delta$ the colatitude. Entrainment by the star is included in the phase term $\Omega\,t$. At the same time the magnetic moment~$\bmu$ points toward a direction depicted by the angles $(\alpha,\beta)$ and given by the unit vector
\begin{equation}
\mathbf{m} = (\sin \alpha \, \cos (\beta+\Omega\,t), \sin \alpha \, \sin (\beta+\Omega\,t), \cos \alpha) .
\end{equation}
The observer line of sight represented by the unit vector~$\nobs$ is by convention located in the~$(xOz)$ plane, forming an angle~$\zeta$ with the spin axis ($z$ axis) thus 
\begin{equation}
\nobs = (\sin \zeta, 0, \cos \zeta) .
\end{equation}
All important parameters are summarized in fig.~\ref{fig:Dipole}.

The magnetic poles are defined by the intersection between the stellar surface (a sphere of radius~$R$) and the magnetic moment axis. Their positions are found following the procedure we now describe. Let a sphere of radius~$R$ be centred at the origin of the reference frame. The intersection between this sphere and the straight line passing through the magnetic dipole moment located at~$M$ along its direction $\mathbf{m}$ is parametrized by a real parameter~$\lambda$ such that $\mathbf{r} = \lambda \, \mathbf{m} + \mathbf{d}$. We look for values of the parameter~$\lambda$ satisfying the relation $||\mathbf{r}|| = R$. This is equivalent to a quadratic equation in $\lambda$ requiring $\lambda^2 + 2 \, \lambda \, \mathbf{m} \cdot \mathbf{d} + d^2 - R^2 = 0$. The discriminant of this equation is equal to $\Delta = 4 \, ( (\mathbf{m} \cdot \mathbf{d})^2 + R^2 - d^2 )$ and always positive since $d<R$. Solutions are therefore always real and equal to
\begin{equation}
\label{eq:lambda}
 \lambda_\pm = -\mathbf{m} \cdot \mathbf{d} \pm \sqrt{(\mathbf{m} \cdot \mathbf{d})^2 + R^2 - d^2}
\end{equation}
from which we deduce the poles at position
\begin{equation}
\label{eq:pole}
\mathbf{r}_\pm = \lambda_\pm \, \mathbf{m} + \mathbf{d} .
\end{equation}
The reference direction for zero polarization angle is arbitrary. We could chose the plane containing the rotation axis and one magnetic pole as a reference but we find it more convenient to define it by the projection of the rotation axis onto the plane of the sky. The unit vector~$\bomega$ about this direction has components $\bomega  = \nobs \wedge \ey = (-\cos\zeta,0,\sin\zeta)$ satisfying $\bomega\cdot\nobs=0$.

%

\subsection{Polarization properties}

We are looking for the polarization angle evolution with respect to time. The polarization plane~$\Pi$ is defined by the magnetic moment axis and the intersection point between the line of sight and the stellar surface. We first note that the polarization plane~$\Pi$ always rotates around the magnetic moment~$\bmu$ because field lines in the pure dipole geometry are always contained in a plane including the location of the magnetic axis and the direction $\bmu$. More generally we will use a sphere of radius $R+h$ where $h$ is the emission altitude because we demonstrate later that contrary to the RVM, our DRVM depends naturally on the altitude~$h$. Let us denote this intersection of $\nobs$ with the sphere of radius $R+h$ by the point~$A$, fig.~\ref{fig:Dipole}. By definition it possesses Cartesian coordinates given by $(R+h)\,(\sin\zeta,0,\cos\zeta)$. The point~$M$ is a second point contained in the polarization plane~$\Pi$. A normal vector~$\npi$ to the polarization plane~$\Pi$, not necessarily normalized, is simply obtained by the cross product $\npi = \bmu \wedge \mathbf{AM}$.

In the special case when $\bmu \parallel \mathbf{AM}$, the polarization plane is ill-defined. The unit vector orthogonal to the plane of the sky~$\varSigma$ is $\nobs$. The direction of linear polarization is given by the intersection between the planes $\Pi$ and $\varSigma$. It is a straight line pointing in the direction of the vector $\np = \npi \wedge \nobs$.

The reference direction is taken to be the projection of the rotation axis along the plane of the sky thus $\bomega$. The cosinus of the polarization angle~$\psi$ is therefore
\begin{equation}
\cos\psi = \frac{\bomega \cdot \np}{||\bomega|| \cdot ||\np||}
\end{equation}
leaving the sign of~$\psi$ undefined. However, it can be determined by the value of $\sin\psi$ thus by the cross product $\bomega \wedge \np$ which is directed along $\nobs$ (remember that $\bomega\cdot\nobs=0$). We arrive at
\begin{equation}
\sin\psi = \frac{(\bomega \wedge \np) \cdot \nobs}{||\bomega|| \cdot ||\np||} .
\end{equation}
Finally, the tangent of the polarization angle reduces to the closed analytical expression
\begin{equation}
\label{eq:tanpsi}
\tan\psi = \frac{(\bomega \wedge \np) \cdot \nobs}{\bomega \cdot \np} .
\end{equation}
There is no need to work with normalized vectors $\bomega$ and $\np$ because their normalization cancels out in the ratio. However, $\nobs$ must be normalized which is the case. Eq.~(\ref{eq:tanpsi}) is the key result of this letter, written in a concise but precise way.
For later convenience, it is appropriate to normalize distances with respect to the stellar radius~$R$. Thus the translation distance is $d=\epsilon\,R$ and the emission height is $h=\eta\,R$. Going through the definition of all the above mentioned vectors, the polarization angle is found explicitly through the substitution
\begin{subequations}
\label{eq:tanpsilong}
\begin{align}
& (\bomega \wedge \np) \cdot \nobs = ( 1 + \eta - \epsilon \, \cos\delta \, \cos\zeta ) \, \sin\alpha \, \sin(\beta+\varphi)\nonumber  \\
& + \epsilon \, \sin\delta \, ( \cos\alpha \, \cos\zeta \, \sin\varphi - \sin\alpha \, \sin\beta \, \sin\zeta) \\
& \bomega \cdot \np = ( 1 + \eta ) \, ( \cos\alpha \, \sin\zeta - \sin\alpha \, \cos\zeta \, \cos(\beta+\varphi) ) \nonumber \\
& + \epsilon \, ( \sin\alpha \, \cos\delta \, \cos (\beta +\varphi ) - \cos\alpha \, \sin\delta \, \cos\varphi) .
\end{align}
\end{subequations}
For a centred dipole we set $\epsilon=0$ and to compare with previous notation we replace $\alpha=\chi$. Eqs.~(\ref{eq:tanpsi})-(\ref{eq:tanpsilong}) then reduce to the RVM model eq.~(\ref{eq:RVM}) with a shift of the origin of the phase~$\beta$ through the replacement $\varphi \rightarrow \beta+\varphi$. This conclusion is easily derived from geometrical considerations.

\section{Polar cap location and polarization}
\label{sec:Resultats}

The parameter space to explore has six dimensions which can be decomposed into four sets, namely
\begin{itemize}
\item two parameters for the magnetic moment $(\alpha, \beta)$.
\item two parameters for the point~$M$ $(d,\delta)$.
\item one parameter for the line of sight~$\zeta$.
\item one parameter for the altitude~$h$.
\end{itemize}
See fig.~\ref{fig:Dipole} for a three dimensional view of the configuration. Being exhaustive by exploring this parameter space for the detailed behaviour of the polarization angle and the magnetic pole excursion would not fit in a reasonable number of pages. Therefore we sorted out a representative sample of judiciously chosen parameters.

\subsection{Magnetic pole geometry}

\begin{figure}
\scalebox{0.9}{\input{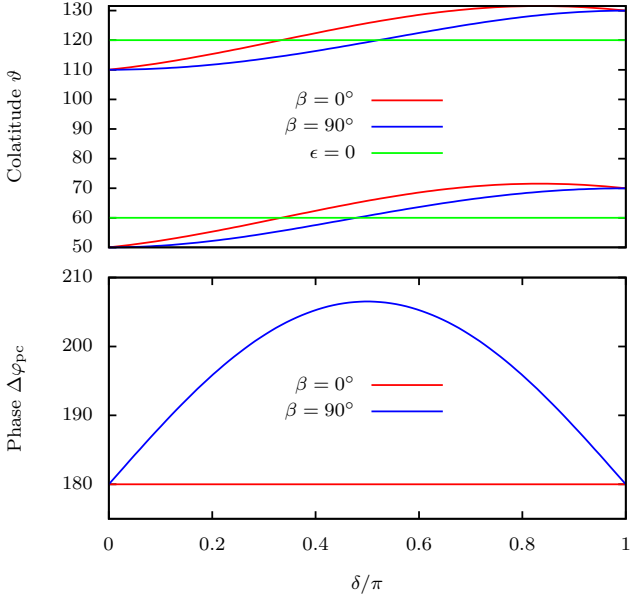}}
\caption{Colatitude of the south and north magnetic pole in the upper panel and phase shift in the lower panel, with the parameters $\alpha=60\degr$, $\beta=0\degr,90\degr$, $\epsilon=0.2$. The reference lines for the centred dipole are marked by ~$\epsilon=0$.}
    \label{fig:Pole}
\end{figure}

The location of the magnetic north and south pole of the dipole, defined as the intersection of the magnetic moment with the neutron star surface, gives insight into the phase lag between both polar caps. The Cartesian coordinates of both poles are given explicitly by eq.~(\ref{eq:pole}). From these coordinates we deduce their colatitude by $\vartheta_\pm = \arccos(z_\pm/R)$. The upper panel of Fig.~\ref{fig:Pole} shows the evolution of the colatitude of both poles with respect to the off-centring angle~$\delta$. Specifically, in this example, the parameters are $\alpha=60\degr$, $\beta=0\degr$ or $90\degr$ and $\epsilon=0.2$. To facilitate comparison with the centred dipole, the constant lines with~$\epsilon=0$ are also shown (the off-centred parameters are therefore irrelevant).
The longitudinal distance between both poles, the so-called phase shift~$\Delta\varphi_{\rm pc}=|\varphi_+-\varphi_-|$ with $\varphi_\pm=\arctan(x_\pm,y_\pm)$ is shown in the lower panel of Fig.~\ref{fig:Pole}. For a centred dipole, the poles are located at $\alpha=60\degr$ and $\upi-\alpha=120\degr$. These are the constant green lines in the upper panel of Fig.~\ref{fig:Pole}. For an off-centred dipole, whatever the value of $\beta$, the poles are travelling around the $\epsilon=0$ case, starting from around $50\degr$ for $\delta=0$ going up to around $130\degr$ for $\delta=\upi$, red and blue lines. However, for $\beta=0\degr$ we notice that both poles follow the same trajectory with a constant phase shift $\Delta\varphi_{\rm pc}=180\degr$, the one expected from a centred dipole. This is confirmed by the red line in the lower panel of Fig.~\ref{fig:Pole}. The most interesting case strongly deviating from the centred dipole is given by $\beta=90\degr$.  The phase shift $\Delta\varphi_{\rm pc}$ experiences a increase from $180\degr$ up to $205\degr$ at $\delta=\upi/2$ followed by a decrease down to $180\degr$ again at $\delta=\upi$, blue line. Consequently, even for a dipole field, if both polar caps are visible, they are not necessarily separated by half a period. An orthogonal rotator with $\alpha=90\degr$ could show phase shift values $\Delta\varphi_{\rm pc}\neq180\degr$ depending on $\beta$,  $\delta$ and $\epsilon$. An off-centred dipole offers much more flexibility and freedom about polar caps relative positions than a centred one but at the expense of introducing more free parameters to adjust. The deviation from $180\degr$ shown by $\Delta\varphi_{\rm pc}$ may be relevant for uncollimated thermal emission like for instance soft X-rays. For radio pulses, usually assumed to be strongly collimated along field lines, computation of the phase lag requires knowledge about the orientation of the lines where photons are emitted. Each pulse is seen in the same phase~$\varphi$ as its corresponding magnetic pole only in the centred dipole. For an off-centred dipole it is no more the case. Radio emission is narrowly collimated along the magnetic axis sticking out from the polar cap. Allowing for the magnetic axis to be misaligned with respect to the position vector~$\mathbf{r}$ of the magnetic pole will lead to pulsation being detected at a different phase~$\varphi$. Indeed, considering only emission straight from the magnetic axis, pulses are observed if and only if $\bmu$ and $\nobs$ are aligned, i.e. $\bmu\wedge\nobs=\mathbf{0}$. Restricting the line of sight inclination angle to $[0,\upi]$, this happens for $\zeta=\alpha$ or $\zeta=\upi-\alpha$ corresponding respectively to phases $\beta+\Omega\,t=0$ and $\beta+\Omega\,t=\upi$. We conclude that when collimation along field lines is taken into account, the phase lag between both pulses is always equal to half a period whatever the geometry because the north pulse is seen at $t_{\rm north}=-\beta/\Omega$ and the south pulse at $t_{\rm south}=(\upi-\beta)/\Omega$ thus $\Delta t=(t_{\rm south}-t_{\rm north}) = P/2$. This is easily understood because we see emission from a straight line going into two opposite directions, requiring a rotation of $180\degr$. Therefore for an off-centred dipole, there is naturally a phase lag between collimated and uncollimated emission emanating from the polar caps. This fact could be useful to constrain the geometry of the dipole.

\subsection{Phase-resolved polarization}

\begin{figure}
\scalebox{0.7}{\input{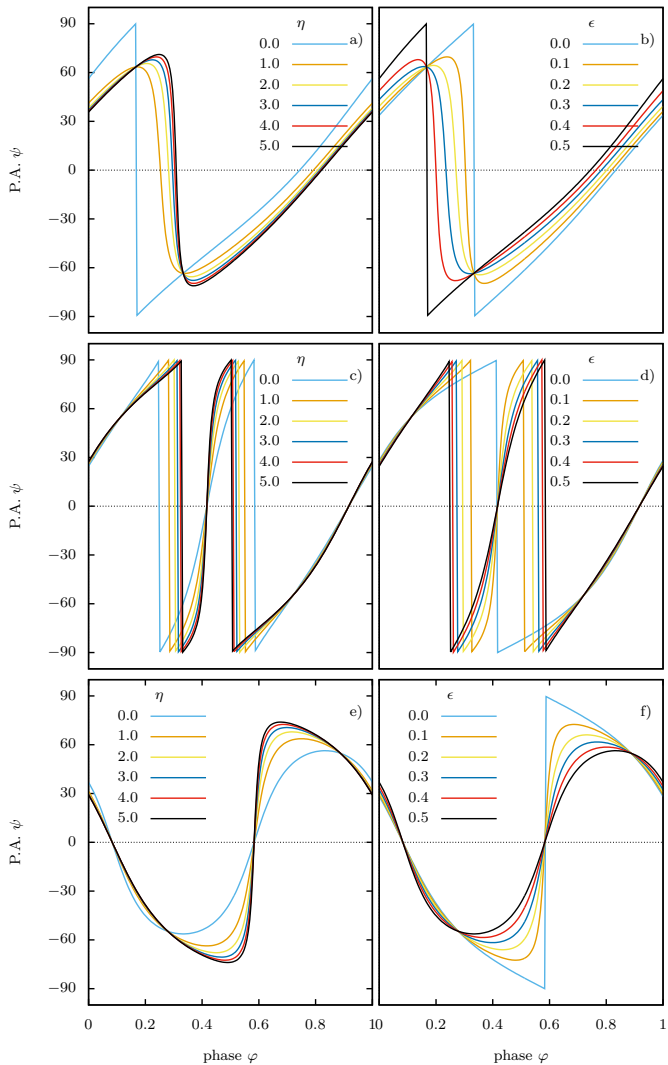}}
\caption{A sample of relevant polarization angle swings for an off-centred dipole. The impact of off-centring and emission altitude are shown separately and quantified respectively with $\epsilon$ and $\eta$. On the left column $\epsilon=0.5$, on the right column $\eta=0$.}
    \label{fig:echantillon}
\end{figure}

Constraining the geometry in the RVM is achieved by phase-resolved polarization measurements. Unfortunately, these observations are not always truly constraining because of large uncertainties in the fits. Nevertheless, polarization is expected to give useful insight into the magnetic field topology in the vicinity of the stellar surface. Implications of off-centring is summarized in Fig.~\ref{fig:echantillon} which depicts a good representative sample of P.A. swings. The corresponding geometry associated to these plots are reported in Table~\ref{tab:echantillon}.
\begin{table}
	\centering
	\caption{The geometry associated to the sample of polarization angle swings as shown in Fig.~\ref{fig:echantillon}.}
	\label{tab:echantillon}
	\begin{tabular}{crrrr} 
		\hline
		plot & $\alpha$ & $\beta$ & $\delta$ & $\zeta$ \\
		\hline
		a), b) & 30\degr & 60\degr & 90\degr & 150\degr \\
		c), d) & 90\degr & 150\degr & 0\degr & 90\degr \\
		e), f) & 60\degr & 30\degr & 0\degr & 120\degr \\
		\hline
	\end{tabular}
\end{table}
The impact of magnetic moment translation is investigation in Fig.~\ref{fig:echantillon}. The influence of the parameter~$\epsilon$ is shown on the right column, panel a), c), e) and the impact of emission height via the parameter~$\eta$ is shown on the left column, panel b), d), f). The reference polarization curves from the RVM are shown in blue solid line on the right column, with $\epsilon=0$. In panel b), the magnetic moment lies in the $(xOy)$ plane with increasing distance $\epsilon=\{0,0.1,0.2,0.3,0.4,0.5\}$. Shifting $\bmu$ to the outer region induces a shift in the polarization angle swing occurring before the one predicted by RVM. Note that except for $\epsilon=\{0,0.5\}$ where a full turn is observed between $[-90\degr,90\degr]$, $\psi$ remains in smaller bounds roughly in the interval $[-60\degr,60\degr]$. The angle $\psi$ rotates slowly counter-clockwise followed by a sharp clockwise swing around the phase $\varphi=0.3$. In panel d), the situation is drastically different, DRVM reaches the bounds $[-90\degr,90\degr]$ twice whereas RVM only once. The P.A. swing is much steeper in DRVM at phase $\varphi=0.4$. The observational signature discrepancies are clearly highlighted in this geometry. In a last set-up, panel f), around phase $\varphi=0.6$, we observe a change of sign in the P.A. swing from DRVM as compared to RVM. But in both cases the P.A. is zero. This contrasts with the interval around phase $\varphi=0.1$ where both swings remain very similar with $\psi$ passing through zero at the same phase. On panel a), c), e) the impact of emission height is investigated in the case $\epsilon=0.5$. In order to facilitate comparison, the blue solid lines on left column are identical to the black solid lines on right column. In panel a), we notice that increasing $\eta$ leads to similar behaviour as going from RVM to an important DRVM but with a phase shift in opposite direction. In panel c), emitting at higher altitudes implies a sharper P.A. gradient around phase $\varphi=0.4$. Panel e) resembles panel f) and shows that larger off-centring is counterbalanced by higher emission heights. As a general trend, around P.A. $\psi=0$, high altitude~$\eta$ implies sharper gradient. 

\section{conclusion}
\label{sec:Discussion}

There is plenty of behaviours to be compared but the best way would be to apply the DRVM to real data from pulsars. This is left for future work. Moreover, deviations from the vacuum static off-centred dipole as presented here are numerous. Rotation leads to stationary vacuum solutions including retarded fields, a generalization of Deutsch solution \citep{deutsch_electromagnetic_1955}. Aberration of light may also distort the polarization angle swing. Plasma screening and currents within the magnetosphere perturb the vacuum solution especially at high altitude. Last but not least, general-relativistic spacetime curvature perturbations on polarization can be substantial. All these effects introduce distortion of the results presented in this letter. We plan to add these corrections as an effort to quantitatively explain the wealth of radio polarization data available so far. Impact of off centring on circular polarization although not mentioned in this work requires also full investigation to even better constrain the DRVM. Our long lasting goal is to quantitatively fit polarization data from pulsars as an indirect diagnostic for their magnetic topology.

\section*{Acknowledgements}

I am grateful to the referee for constructive comments and suggestions. This work has been supported by the French National Research Agency (ANR) through the grant No. ANR-13-JS05-0003-01 (project EMPERE).





\begin{thebibliography}{}
\makeatletter
\relax
\def\mn@urlcharsother{\let\do\@makeother \do\$\do\&\do\#\do\^\do\_\do\%\do\~}
\def\mn@doi{\begingroup\mn@urlcharsother \@ifnextchar [ {\mn@doi@}
  {\mn@doi@[]}}
\def\mn@doi@[#1]#2{\def\@tempa{#1}\ifx\@tempa\@empty \href
  {http://dx.doi.org/#2} {doi:#2}\else \href {http://dx.doi.org/#2} {#1}\fi
  \endgroup}
\def\mn@eprint#1#2{\mn@eprint@#1:#2::\@nil}
\def\mn@eprint@arXiv#1{\href {http://arxiv.org/abs/#1} {{\tt arXiv:#1}}}
\def\mn@eprint@dblp#1{\href {http://dblp.uni-trier.de/rec/bibtex/#1.xml}
  {dblp:#1}}
\def\mn@eprint@#1:#2:#3:#4\@nil{\def\@tempa {#1}\def\@tempb {#2}\def\@tempc
  {#3}\ifx \@tempc \@empty \let \@tempc \@tempb \let \@tempb \@tempa \fi \ifx
  \@tempb \@empty \def\@tempb {arXiv}\fi \@ifundefined
  {mn@eprint@\@tempb}{\@tempb:\@tempc}{\expandafter \expandafter \csname
  mn@eprint@\@tempb\endcsname \expandafter{\@tempc}}}

\bibitem[\protect\citeauthoryear{Backer}{Backer}{1976}]{backer_pulsar_1976}
Backer D.~C.,  1976, \mn@doi [{\textbackslash}apj] {10.1086/154788}, 209, 895

\bibitem[\protect\citeauthoryear{Blaskiewicz, Cordes  \& Wasserman}{Blaskiewicz
  et~al.}{1991}]{blaskiewicz_relativistic_1991}
Blaskiewicz M.,  Cordes J.~M.,   Wasserman I.,  1991, \mn@doi [ApJ]
  {10.1086/169850}, 370, 643

\bibitem[\protect\citeauthoryear{Borra}{Borra}{1974}]{borra_orientation_1974}
Borra E.~F.,  1974, \mn@doi [The Astrophysical Journal] {10.1086/152624}, 187,
  271

\bibitem[\protect\citeauthoryear{Chung \& Melatos}{Chung \&
  Melatos}{2011}]{chung_stokes_2011}
Chung C. T.~Y.,  Melatos A.,  2011, \mn@doi [MNRAS]
  {10.1111/j.1365-2966.2010.17858.x}, 411, 2471

\bibitem[\protect\citeauthoryear{Craig}{Craig}{2014}]{craig_tackling_2014}
Craig H.~A.,  2014, \mn@doi [ApJ] {10.1088/0004-637X/790/2/102}, 790, 102

\bibitem[\protect\citeauthoryear{Craig \& Romani}{Craig \&
  Romani}{2012}]{craig_altitude_2012}
Craig H.~A.,  Romani R.~W.,  2012, \mn@doi [The Astrophysical Journal]
  {10.1088/0004-637X/755/2/137}, 755, 137

\bibitem[\protect\citeauthoryear{Deutsch}{Deutsch}{1955}]{deutsch_electromagnetic_1955}
Deutsch A.~J.,  1955, Annales d'Astrophysique, 18, 1

\bibitem[\protect\citeauthoryear{Gil \& Mitra}{Gil \&
  Mitra}{2001}]{gil_vacuum_2001}
Gil J.,  Mitra D.,  2001, \mn@doi [{\textbackslash}apj] {10.1086/319714}, 550,
  383

\bibitem[\protect\citeauthoryear{Gil, Melikidze  \& Mitra}{Gil
  et~al.}{2002}]{gil_modelling_2002}
Gil J.~A.,  Melikidze G.~I.,   Mitra D.,  2002, \mn@doi [{\textbackslash}aap]
  {10.1051/0004-6361:20020473}, 388, 235

\bibitem[\protect\citeauthoryear{Hibschman \& Arons}{Hibschman \&
  Arons}{2001}]{hibschman_polarization_2001}
Hibschman J.~A.,  Arons J.,  2001, \mn@doi [{\textbackslash}apj]
  {10.1086/318224}, 546, 382

\bibitem[\protect\citeauthoryear{Izvekova, Malov  \& Malofeev}{Izvekova
  et~al.}{1977}]{izvekova_applicability_1977}
Izvekova V.~A.,  Malov I.~F.,   Malofeev V.~M.,  1977, Soviet Astronomy
  Letters, 3, 237

\bibitem[\protect\citeauthoryear{Komesaroff}{Komesaroff}{1970}]{komesaroff_possible_1970}
Komesaroff M.~M.,  1970, \mn@doi [{\textbackslash}nat] {10.1038/225612a0}, 225,
  612

\bibitem[\protect\citeauthoryear{Komesaroff, Morris  \& Cooke}{Komesaroff
  et~al.}{1970}]{komesaroff_linear_1970}
Komesaroff M.~M.,  Morris D.,   Cooke D.~J.,  1970, {\textbackslash}aplett, 5,
  37

\bibitem[\protect\citeauthoryear{Landstreet}{Landstreet}{1970}]{landstreet_orientation_1970}
Landstreet J.~D.,  1970, \mn@doi [The Astrophysical Journal] {10.1086/150377},
  159, 1001

\bibitem[\protect\citeauthoryear{Lyne \& Manchester}{Lyne \&
  Manchester}{1988}]{lyne_shape_1988}
Lyne A.~G.,  Manchester R.~N.,  1988, {\textbackslash}mnras, 234, 477

\bibitem[\protect\citeauthoryear{Phillips}{Phillips}{1992}]{phillips_radio_1992}
Phillips J.~A.,  1992, \mn@doi [{\textbackslash}apj] {10.1086/170936}, 385, 282

\bibitem[\protect\citeauthoryear{P\'etri}{P\'etri}{2016}]{petri_radiation_2016}
P\'etri J.,  2016, \mn@doi [MNRAS] {10.1093/mnras/stw2050}, p. stw2050

\bibitem[\protect\citeauthoryear{Radhakrishnan}{Radhakrishnan}{1992}]{radhakrishnan_polarization_1992}
Radhakrishnan V.,  1992, in Hankins T.~H.,  Rankin J.~M.,   Gil J.~A.,  eds,
  {IAU} {Colloq}. 128: {Magnetospheric} {Structure} and {Emission} {Mechanics}
  of {Radio} {Pulsars}. p.~367

\bibitem[\protect\citeauthoryear{Radhakrishnan \& Cooke}{Radhakrishnan \&
  Cooke}{1969}]{radhakrishnan_magnetic_1969}
Radhakrishnan V.,  Cooke D.~J.,  1969, {\textbackslash}aplett, 3, 225

\bibitem[\protect\citeauthoryear{Smith}{Smith}{1970}]{smith_beaming_1970}
Smith F.~G.,  1970, {\textbackslash}mnras, 149, 1

\bibitem[\protect\citeauthoryear{Stift}{Stift}{1974}]{stift_decentred_1974}
Stift M.~J.,  1974, \mn@doi [Monthly Notices of the Royal Astronomical Society]
  {10.1093/mnras/169.3.471}, 169, 471

\bibitem[\protect\citeauthoryear{Xilouris, Kramer, Jessner, von Hoensbroech,
  Lorimer, Wielebinski, Wolszczan  \& Camilo}{Xilouris
  et~al.}{1998}]{xilouris_characteristics_1998}
Xilouris K.~M.,  Kramer M.,  Jessner A.,  von Hoensbroech A.,  Lorimer D.~R.,
  Wielebinski R.,  Wolszczan A.,   Camilo F.,  1998, \mn@doi [The Astrophysical
  Journal] {10.1086/305791}, 501, 286

\bibitem[\protect\citeauthoryear{Yan et~al.,}{Yan
  et~al.}{2011}]{yan_polarization_2011}
Yan W.~M.,  et~al., 2011, \mn@doi [MNRAS] {10.1111/j.1365-2966.2011.18522.x},
  414, 2087

\bibitem[\protect\citeauthoryear{Zhang, Harding  \& Muslimov}{Zhang
  et~al.}{2000}]{zhang_radio_2000}
Zhang B.,  Harding A.~K.,   Muslimov A.~G.,  2000, \mn@doi [The Astrophysical
  Journal Letters] {10.1086/312542}, 531, L135

\makeatother
\end{thebibliography}




%


\bsp	
\label{lastpage}
\end{document}